\documentclass
[aps,prd,preprintnumbers,titlepage,nofootinbib,twocolumn]{revtex4-2}%
\usepackage[T1]{fontenc}
\usepackage[latin9]{inputenc}
\usepackage{amsmath}
\usepackage{amssymb}
\usepackage{stmaryrd}
\usepackage{stackrel}
\usepackage{graphicx}
\usepackage[T1]{fontenc}
\usepackage[latin9]{inputenc}
\usepackage{amsmath}
\usepackage{amsfonts,amssymb}
\usepackage{latexsym}
\usepackage{epsfig}
\usepackage{graphicx}
\usepackage{tikz}
\usepackage[colorlinks,linkcolor=greatblue,anchorcolor=blue,citecolor=greatred]%
{hyperref}
\usepackage{graphicx}
\usepackage{dcolumn}
\usepackage{bm}
\usepackage{babel}
\usepackage{amsfonts}%
\makeatletter
\makeatletter
\def\fnum@figure{\text{FIG.~\thefigure}}
\definecolor{greatblue}{RGB}{40,120,181}
\definecolor{greatred}{RGB}{200,36,35}
\makeatother
\begin{document}
\preprint{CTP-SCU/20250017}
\title{Coexistence of Spectrally Stable and Unstable Modes in Black Hole Ringdowns}
\author{Peng Wang$^{a}$}
\email{pengw@scu.edu.cn}
\author{Tianshu Wu$^{b,a}$}
\email{wuts25@mails.tsinghua.edu.cn}
\affiliation{$^{a}$College of Physics, Sichuan University, Chengdu, 610065, China}
\affiliation{$^{b}$Department of Astronomy, Tsinghua University, Beijing 100084, China}

\begin{abstract}
Recent studies have shown that a secondary potential barrier, forming a
potential well outside the event horizon, can destabilize the Quasinormal Mode
(QNM) spectrum of black holes. We find that spectral instability may persist
even after the potential well vanishes, giving rise to a distinct family of
spectrally unstable QNMs that differ from the spectrally stable modes
localized near the potential peak and associated with the photon sphere.
Nevertheless, time-domain simulations reveal that early-time ringdown
waveforms remain dominated by stable modes, while unstable modes have only a
subdominant contribution. These results highlight the robustness of black hole
spectroscopy, as the observable ringdown signal is primarily governed by the
most stable QNMs.

\end{abstract}
\maketitle

\section*{Introduction}

The direct detections of gravitational waves by the LIGO-Virgo-KAGRA
Collaboration have opened a new era of Black Hole (BH) physics
\cite{LIGOScientific:2016aoc,LIGOScientific:2017vwq,LIGOScientific:2018dkp,LIGOScientific:2020tif,LIGOScientific:2025hdt}%
, allowing us to probe the nature of strong gravity through BH spectroscopy,
which measures the characteristic Quasinormal Modes (QNMs) emitted by a
perturbed remnant BH
\cite{Vishveshwara:1970cc,Teukolsky:1973ha,Chandrasekhar:1975zza,Leaver:1985ax,Chandrasekhar:1985kt,Leaver:1986gd,Kokkotas:1999bd,Nollert:1999ji,Dreyer:2003bv,Berti:2009kk,Konoplya:2011qq,Baibhav:2017jhs,Berti:2025hly}%
. Each QNM encodes information about the mass, spin, and possible deviations
from the Kerr solution, providing a powerful test of general relativity in the
nonlinear regime
\cite{Carter:1971zc,Robinson:1975bv,LIGOScientific:2016lio,Price:2017cjr,Carullo:2019flw,Giesler:2019uxc,Cardoso:2019rvt,Bhagwat:2019dtm,CalderonBustillo:2020rmh,Isi:2021iql,Capano:2021etf,Ma:2023cwe,Baibhav:2023clw,LIGOScientific:2025obp}%
. The success of the BH spectroscopy program crucially relies on the
assumption that the QNM spectrum is physically robust and stable under small
perturbations to the underlying spacetime.

Recent investigations, however, have revealed intriguing signs that this
assumption may not always hold. Several studies have shown that when a small
\textquotedblleft bump\textquotedblright\ is introduced into the curvature
potential governing BH perturbations, the resulting QNM spectrum can undergo
drastic rearrangements or migrations, signaling spectral instability
\cite{Cheung:2021bol,Courty:2023rxk,Cardoso:2024mrw,Ianniccari:2024ysv,Yang:2024vor,Boyanov:2024fgc,Laeuger:2025zgb,Shen:2025yiy,Mai:2025cva,MalatoCorrea:2025iuc,Wu:2025sbq}%
. Complementary analyses based on the pseudospectrum have further demonstrated
that the QNM operator is highly non-normal, so that even tiny perturbations in
the potential can induce large spectral shifts
\cite{Jaramillo:2020tuu,Destounis:2021lum,Jaramillo:2021tmt,Jaramillo:2022kuv,Sarkar:2023rhp,Arean:2023ejh,Cao:2024oud,Chen:2024mon,Luo:2024dxl,Cao:2024sot,Siqueira:2025lww,Cai:2025irl,Cao:2025qws}%
. These developments have raised important questions about the mathematical
stability and physical interpretability of QNMs, which form the theoretical
foundation of BH spectroscopy.

To assess the significance of this issue, it is essential to examine not only
the frequency-domain spectrum but also the time-domain response of the system.
Indeed, time-domain studies have provided reassuring evidence that the
ringdown waveforms themselves remain stably perturbed even in cases where the
QNM spectrum appears unstable
\cite{Barausse:2014tra,Jaramillo:2021tmt,Berti:2022xfj,Spieksma:2024voy,Oshita:2025ibu}%
. Moreover, alternative frequency-based characterizations of BH perturbations,
such as greybody factors and ringdown filters, have been shown to remain
robust against the same types of perturbations that trigger spectral
instability in the QNM spectrum
\cite{Kyutoku:2022gbr,Ma:2022wpv,Oshita:2023cjz,Torres:2023nqg,Guo:2023nkd,Okabayashi:2024qbz,Rosato:2024arw,Oshita:2024fzf,Li:2025ljb,Xie:2025jbr}%
. These findings suggest that spectral instability may primarily reflect the
mathematical sensitivity of the QNM eigenvalue problem rather than a physical
instability in the observable ringdown signal.

Despite these insights, the physical mechanism underlying spectral instability
remains to be clarified. From a physical perspective, spectral instability
often arises in systems where the effective potential develops a potential
well induced by the added bump. The appearance of this well can give rise to a
new family of long-lived modes that may eventually overtake the fundamental
QNM, leading to apparent spectral instability. One might thus expect that when
the potential well disappears, the QNM spectrum should recover its spectral
stability. However, as we will show, the situation is more subtle: even when
the potential well vanishes, a residual scale inherited from the potential
well can still produce a family of spectrally unstable QNMs.

This observation naturally raises a deeper question: if both spectrally stable
and unstable QNM families coexist within the same system, does the presence of
spectrally unstable modes necessarily lead to observable instabilities in the
time-domain ringdown signal, thereby threatening the reliability of BH
spectroscopy? In this work, we explore this question in a toy yet physically
viable model: a test scalar field propagating on a static, spherically
symmetric BH background in the Einstein-Maxwell-scalar (EMS) model
\cite{Herdeiro:2018wub,Gan:2021xdl,Zhang:2021nnn,Guo:2024cts}. By combining
frequency- and time-domain analyses, we demonstrate how spectral (in)stability
manifests in such systems and assess its physical impact on ringdown signals.

\section*{Setup}

In this work, we exam the QNM spectrum of a test scalar field in the static,
spherically symmetric BH background in an EMS model. The EMS model describes a
scalar field $\phi$ minimally coupled to the metric field and non-minimally
coupled to electromagnetic field $A_{\mu}$, with the action
\begin{equation}
S=\int d^{4}x\sqrt{-g}\left[  R-2\left(  \partial\phi\right)  ^{2}%
-e^{\alpha\phi^{2}}F^{\mu\nu}F_{\mu\nu}\right]  . \label{eq:Action}%
\end{equation}
Here, $F_{\mu\nu}=\partial_{\mu}A_{\nu}-\partial_{\nu}A_{\mu}$ is the
electromagnetic field strength tensor, and $e^{\alpha\phi^{2}}$ is the
coupling function between $\phi$ and $A_{\mu}$. Interestingly, adopting the
spherically symmetric and asymptotically flat BH ansatz%
\begin{equation}
ds^{2}=g^{\mu\nu}dx_{\mu}dx_{\nu}=-N\left(  r\right)  e^{-2\delta\left(
r\right)  }dt^{2}+\frac{dr^{2}}{N\left(  r\right)  }+r^{2}d\Omega,
\label{eq:metric}%
\end{equation}
hairy BH solutions with a non-trivial scalar field profile can be numerically
constructed via the shooting or spectrum methods
\cite{Herdeiro:2018wub,Guo:2023mda}. We briefly discuss in the Supplemental
Material how to compute BH solutions using the spectral method.

For simplicity, we consider the perturbation of a massless neutral scalar
field $\Psi$ in the hairy BH background, which is governed by
\begin{equation}
\oblong\Psi=\frac{1}{\sqrt{-g}}\partial_{\mu}\left(  g^{\mu\nu}\sqrt
{-g}\partial_{\nu}\Psi\right)  =0. \label{eq:eom}%
\end{equation}
By decomposing the perturbation into spherical harmonics $Y_{lm}\left(
\theta,\varphi\right)  $, the field can be expressed as $\Psi=r^{-1}\sum
_{l,m}\psi\left(  t,r\right)  Y_{lm}\left(  \theta,\varphi\right)  $. Eq.
$\left(  \ref{eq:eom}\right)  $ then reduces to the Regge-Wheeler-Zerilli type
wave equation,%
\begin{equation}
\left(  -\frac{\partial^{2}}{\partial t^{2}}+\frac{\partial^{2}}{\partial
x^{2}}-V_{\text{eff}}\left(  x\right)  \right)  \psi\left(  t,x\right)  =0,
\label{eq:t-x eq}%
\end{equation}
where the tortoise coordinate $x$ is determined by $dx/dr\equiv e^{\delta
\left(  r\right)  }N^{-1}\left(  r\right)  $, and the effective potential is
given by
\begin{equation}
V_{\text{eff}}\left(  x\right)  =\frac{e^{-2\delta}N}{r^{2}}\left[  l\left(
l+1\right)  +1-N-\frac{Q^{2}}{r^{2}e^{\alpha\phi^{2}}}\right]  ,
\label{eq:veff}%
\end{equation}
with $Q$ being BH charge $Q$. To analyze the QNM spectrum, one typical
preforms a Fourier transform $\psi\left(  t,x\right)  =\int d\omega\tilde
{\psi}\left(  \omega,x\right)  e^{-i\omega t}$, that reduces\ the wave
equation to a master equation for QNMs,%
\begin{equation}
\left(  \frac{d^{2}}{dx^{2}}+\omega^{2}-V_{\text{eff}}\left(  x\right)
\right)  \tilde{\psi}\left(  \omega,x\right)  =0. \label{eq:PertubaEq}%
\end{equation}
Subsequently, QNMs are determined by imposing purely ingoing waves at the
horizon and purely outgoing waves at infinity, yielding a discrete set of QNMs
with complex frequencies $\omega_{n}=\omega_{R}+i\omega_{I}$, where
$n=0,1,2...$ denotes the overtone number.

Nevertheless, computing QNMs---especially overtones---of BHs is numerically
challenging when the background BH solution is constructed numerically. To
minimize numerical errors, we employ spectral methods with identical
resolutions to compute both the background BH solution and its associated
QNMs. The metric functions are evaluated directly at the spectral grid points,
and these values are subsequently used to construct the coefficients of the
QNM master equation $\left(  \ref{eq:PertubaEq}\right)  $ on the same grid,
thereby ensuring consistency and reducing interpolation errors. Alternatively,
one can evolve the wave equation $\left(  \ref{eq:t-x eq}\right)  $ in time
and extract QNMs from the resulting waveform. The technical details of our
numerical implementation are presented in the Supplemental Material.

\section*{QNM Spectrum}

\begin{figure*}[ptb]
\includegraphics[width=0.95\linewidth]
{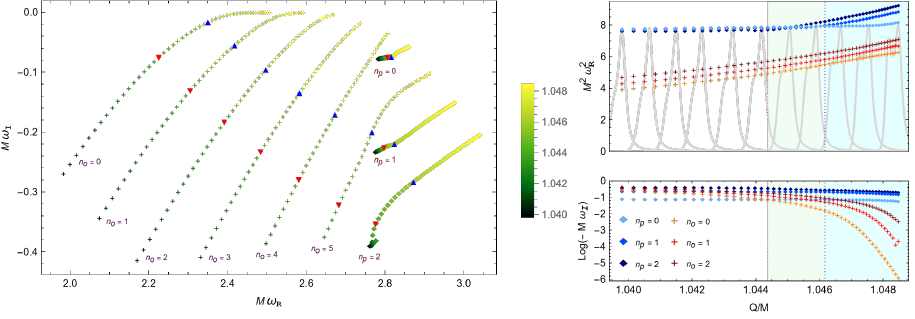}\caption{QNM spectrum of the scalar field with $l=10$ for hairy
BHs with varying charge-to-mass ratio $Q/M$. Diamonds denote the peak modes
$\omega_{n}^{(p)}$, indexed by $n_{p}$, while plus signs indicate the off-peak
modes $\omega_{n}^{(o)}$, indexed by $n_{o}$. \textbf{Left:}\textit{
}Trajectories of QNM frequencies in the complex plane, illustrating how the
modes migrate as $Q/M$ varies. The color bar indicates $Q/M$. Diamonds denote
the peak modes, indexed by $n_{p}$, while plus signs indicate the off-peak
modes, indexed by $n_{o}$. Blue triangles mark the transition points separating the single-peaked
and double-peaked regimes of the effective potential $V_{\text{eff}}$; the
portions of the trajectories below and to the left of the triangles correspond
to the single-peaked regime. In this regime, the off-peak modes exhibit much
larger frequency migrations than the peak modes, indicating that the latter
are more spectrally stable under variations in $Q/M$. \textbf{Upper right}:
Square of the real part of the QNM frequency, $\omega_{R}^{2}$, as a function
of $Q/M$, together with representative profiles of $V_{\text{eff}}$. The
highest peak of $V_{\text{eff}}$ corresponds to each selected $Q/M$. The
cyan-shaded region indicates the double-peaked regime. The peak modes are
associated with oscillations near the potential peak (photon-sphere modes) in
the single-peaked regime, whereas the low-lying off-peak modes correspond to
states trapped within the potential valley in the double-peaked regime.
\textbf{Lower right}: Imaginary part of the QNM frequencies as a function of
$Q/M$. A fundamental-mode overtaking occurs between the peak and off-peak
families. The fundamental QNM is identified as the slowest-decaying mode among
peak and off-peak families in the white- and green-shaded regions,
respectively. Red triangles mark the overtaking event in the complex-frequency
plane.}%
\label{qnnl10}%
\end{figure*}

\begin{figure*}[ptb]
\includegraphics[width=0.95\linewidth]{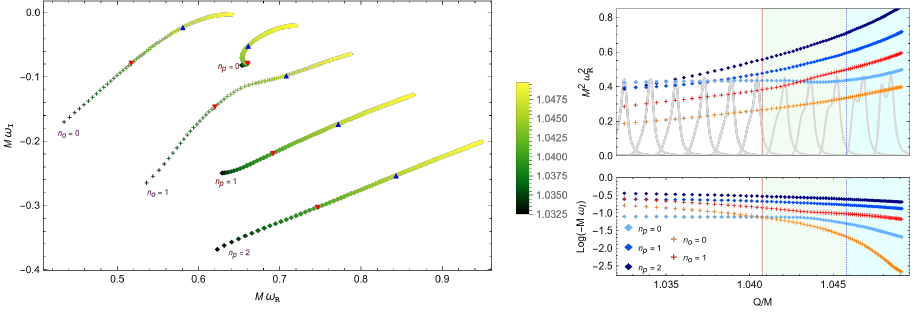}\caption{QNM spectrum of the
scalar field with $l=2$ for hairy BHs with varying charge-to-mass ratio $Q/M$.
Compared with the $l=10$ case, the $l=2$ spectrum exhibits a much stronger
influence from the remnant of the double-peaked structure within the
single-peaked regime. This enhanced residual effect renders the $n_{p}=2$
overtone of the peak-mode family spectrally unstable, while the $n_{p}=0$ and
$n_{p}=1$ modes remain spectrally stable provided $Q/M$ stays sufficiently far
from the double-peaked regime.}%
\label{qnnl2}%
\end{figure*}

For hairy BH backgrounds with $\alpha=0.8$ and varying charge-to-mass ratio
$Q/M$, we compute the QNMs of the scalar field $\Psi$ for $l=2$ and $l=10$, as
shown in Figs. \ref{qnnl10} and \ref{qnnl2}, respectively. The left panels
display the trajectories of the first few QNM frequencies in the complex plane
as $Q/M$ varies, illustrating how the QNM spectrum migrates with respect to
$Q/M$. The upper-right panels show the square of the real part of the QNM
frequency, $\omega_{R}^{2}$, as a function of $Q/M$, together with the
effective potential $V_{\text{eff}}$ for several representative values of
$Q/M$. The relative position of $\omega_{R}^{2}${} with respect to
$V_{\text{eff}}$ provides insights into the physical origin of the QNMs. For
instance, modes with $\omega_{R}^{2}$ near a potential peak or within a
potential valley correspond, respectively, to oscillations around the photon
sphere and to quasi-bound states trapped inside the valley
\cite{Cardoso:2008bp,Guo:2021bcw,Guo:2021enm,Guo:2022umh}.

Interestingly, unlike Schwarzschild or Reissner-Nordstr\"{o}m BHs, the
effective potential $V_{\text{eff}}$ of hairy BHs can develop a double-peaked
structure when $Q/M$ is sufficiently large. The portion of the frequency
trajectories lying beyond (upper-right of) blue triangles in the left panels,
along with the QNMs in the cyan-shaded regions of the right panels, correspond
to this double-peaked regime. Moreover, the upper-right panels reveal that
$V_{\text{eff}}$ transitions from a double-peaked to a single-peaked profile
as $Q/M$ decreases.

We first examine the QNM spectrum of the scalar field with $l=10$, as shown in
Fig. \ref{qnnl10}. Tracing the QNM frequency trajectories from the upper-right
corner, we observe that the spectrum bifurcates into two distinct families
once the effective potential $V_{\text{eff}}${} becomes single-peaked. By
analyzing the behavior of $\omega_{R}^{2}$, these two families can be
identified as follows:

\begin{itemize}
\item \textbf{Peak modes}: These modes are localized near the potential peak
and are associated with the photon sphere situated near that peak. In the
eikonal limit, the real part of the QNM frequency approximately equals
$\sqrt{V_{\text{eff}}}$ at the peak, which also determines the angular
velocity of photons on the unstable circular orbit at the photon sphere
\cite{Cardoso:2008bp}. The modes are indexed by $n_{p}$ according to their
decay rates, starting from the most slowly decaying mode $\left(
n_{p}=0\right)  $. This family resembles the QNMs found in Schwarzschild and
Kerr BHs and is represented by diamonds in Fig. \ref{qnnl10}.

\item \textbf{Off-peak modes}: These modes are found at an appreciable
distance from the potential peak and are marked by plus signs in Fig.
\ref{qnnl10}. Unlike the peak modes, they cannot be attributed to photon
spheres. Remarkably, as $Q/M$ increases and $V_{\text{eff}}$ develops a
double-peaked structure, the low-lying modes evolve into long-lived modes
trapped within the potential valley between the two peaks. This behavior
suggests that the off-peak modes in the single-peaked regime retain a residual
imprint of the valley configuration present in the double-peaked
regime$^{\ref{ft:1}}$ \footnotetext[1]{\label{ft:1} While this work was in
preparation, \cite{Yang:2025hqk} appeared on arXiv, where the authors studied
a hairy Schwarzschild black hole admitting a double-peaked structure.
Interestingly, near the boundary between the single-peaked and double-peaked
regimes, they observed diffraction-trapped states, which share the same
physical origin as the off-peak modes identified in our study.}. The index
$n_{o}$ is used to label these modes in order of their decay rates, from the
slowest $\left(  n_{o}=0\right)  $ to the fastest decaying.
\end{itemize}

Remarkably, in the single-peaked regime, the off-peak modes are noticeably
more sensitive to variations in $Q/M$ than the peak modes, as indicated by
their larger frequency migration distances in the complex plane. This
observation is further supported by the upper-right and lower-right panels of
Fig. \ref{qnnl10}, which show that both the real and imaginary parts of the
peak-mode frequencies exhibit much weaker spectral shifts compared to the
pronounced changes of the off-peak modes. Moreover, the migration rate
$\delta\omega$, defined in the Supplemental Material, confirms that
$\delta\omega$ of the off-peak modes is significantly higher than that of the
peak modes. It is also noteworthy that, near and within the double-peaked{}
regime, the fundamental peak mode $\left(  n_{p}=0\right)  $ remains
spectrally stable, whereas its overtones become increasingly unstable.

Furthermore, as $Q/M$ increases, the imaginary part of the off-peak modes
decreases at a rate substantially faster than that of the peak modes. This
contrasting behavior results in a level crossing, during which the fundamental
QNM---initially identified as the $n_{p}=0$ peak mode---is overtaken by the
$n_{o}=0$ off-peak mode. The QNMs at the point of this overtaking are
indicated by red triangles in the complex frequency plane. Notably, this
transition is accompanied by a clear discontinuity in the real part of the
fundamental QNM frequency and by the onset of destabilization of the
fundamental mode itself.

Fig. \ref{qnnl2} illustrates the spectral migration of the QNM spectrum as
$Q/M$ varies for the $l=2$ case. Similar to the $l=10$ case discussed earlier,
the spectrum in the single-peaked regime consists of two distinct families:
the peak modes and the off-peak modes, between which a fundamental QNM
overtaking occurs. However, a notable difference arises for $l=2$: the
residual influence of the potential valley, a remnant of the double-peaked
structure, exerts a much stronger effect on the QNMs in the single-peaked
regime. This enhanced influence renders the $n_{p}=2$ overtone of the
peak-mode family spectrally unstable, resembling the behavior of the off-peak
modes. In contrast, the $n_{p}=0$ and $n_{p}=1$ peak modes remain spectrally
stable, provided that $Q/M$ stays sufficiently far from the range associated
with the double-peaked $V_{\text{eff}}$. Moreover, in the double-peaked
regime, the potential valley is considerably shallower in the $l=2$ case, so
that only a single off-peak mode, $n_{o}=0$, forms a quasi-bound state trapped
within the valley.

\section*{QNM Extraction}

In this section, we analyze QNM extraction from time-domain waveforms,
focusing on how the spectral (in)stability of QNM modes influences the
extraction results. Specifically, we consider two representative cases:
$Q/M=1.04436$ with $l=10$ and $Q/M=1.03507$ with $l=2$. In the former case,
the peak modes display pronounced spectral stability relative to the off-peak
modes and correspond to the fundamental QNM overtaking, where the fundamental
peak and off-peak modes share the same imaginary part. In the latter case, the
first two modes of the peak-mode family remain substantially more spectrally
stable than their counterparts in the off-peak family.

To obtain the scalar field waveforms, we numerically solve the wave equation
$\left(  \ref{eq:t-x eq}\right)  $ in the time domain, using a Gaussian pulse
located near the potential peak as the initial data. The numerical integration
procedure and the resulting waveforms are presented in the Supplemental
Material. At late times, the waveform decays as a linear combination of
exponentially damped sinusoids, whose frequencies and damping rates can be
extracted by fitting the signal with an $N$-mode template,%
\begin{equation}
\psi\left(  t\right)  =\mathrm{Re}{\sum_{n=0}^{N-1}}A_{n}e^{-i\left(
\omega_{n}t-\phi_{n}\right)  },
\end{equation}
where $A_{n}$ and $\phi_{n}$ denote the amplitude and phase of the $n$-th
mode, respectively.

We employ two fitting schemes for QNM extraction: (i) Strong (agnostic) fit
model: All parameters $\left\{  A_{n},\phi_{n},\omega_{n}\right\}  $ are
treated as free, and we use the NonlinearModelFit function in Mathematica to
determine them. The extracted QNM frequencies can serve as an independent
cross-check of the frequency-domain results. (ii) Weak fit model: The
frequencies are fixed, and the notation $\left(  N_{p},N_{o}\right)  $
specifies the number of modes selected from the peak and off-peak families,
respectively. Only $\left\{  A_{n},\phi_{n}\right\}  $ are treated as fitting
parameters, and the fitting is performed using the LinearModelFit function in
Mathematica. The resulting amplitudes quantify the relative contributions of
each QNM to the time-domain signal. Both fitting models are applied over the
time window $\left[  t_{0},t_{peak}+300M\right]  $, where the start time
$t_{0}$ varies within $\left[  t_{peak},t_{peak}+300M\right]  $, and
$t_{peak}$ denotes the time corresponding to the waveform's maximum amplitude.
If the waveform accurately follows the QNM model, a stable plateau region
emerges in which the fitted parameters remain nearly constant across a range
of start times \cite{Baibhav:2023clw,Takahashi:2023tkb}. More details
regarding the weak fit model are provided in the Supplementary Material.

\begin{figure*}[ptb]
\includegraphics[width=0.95\linewidth]{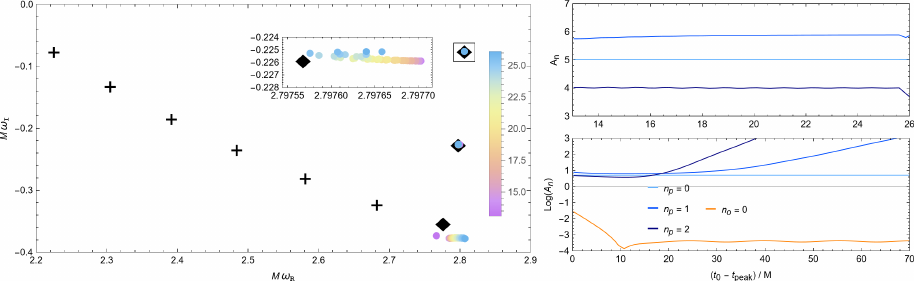}\caption{QNM
fitting results for $Q/M=1.04436$ with $l=10$. The waveform is fitted using
the strong fit model with three QNMs and the weak fit model with $\left(
N_{p},N_{o}\right)  =\left(  3,1\right)  $. \textbf{Left}: QNM frequencies
extracted from the strong fit model in the complex plane for varying start
times, represented by shaded dots. The color bar indicates the start time.
Diamonds and plus signs denote the frequencies of the peak and off-peak
families computed in the frequency domain, respectively. Only the peak modes
can be reliably extracted. \textbf{Upper right}: Extracted amplitudes $A_{n}$
from the strong fit model as functions of the start time $t_{0}$.
\textbf{Lower right}: Amplitudes obtained from the weak fit model including
the first three peak modes and one off-peak mode. The off-peak mode amplitude
is approximately $\mathcal{O}\left(  10^{-4}\right)  $ smaller than those of
the peak modes, demonstrating that the off-peak contribution to the
time-domain waveform is negligible.}%
\label{fig: fitl10}%
\end{figure*}

Fig. \ref{fig: fitl10} presents the fitting results for the case $Q/M=1.04436$
with $l=10$, where the waveform is analyzed using the strong fit model with
three QNMs and the weak fit model with $\left(  N_{p},N_{o}\right)  =\left(
3,1\right)  $. The left panel shows the QNM frequencies extracted from the
strong fit model in the complex plane, with dots of varying shades
representing results obtained at different start times. For reference,
diamonds and plus signs indicate the frequency-domain QNMs belonging to the
peak and off-peak families, respectively. Within $13\lesssim\left(
t_{0}-t_{peak}\right)  /M\lesssim26$, the first three peak modes are
successfully extracted, and their frequencies exhibit good agreement with the
frequency-domain results. Furthermore, as shown in the upper-right panel, the
extracted amplitudes display a clear plateau as functions of $t_{0}$,
providing additional confirmation of the robustness of the fitting.

However, the strong fit model fails to extract the off-peak modes from the
waveform, indicating that the peak modes dominate the QNM contribution to the
time-domain signal. To assess the influence of the off-peak modes, we apply
the weak fit model using the $n_{0}=0$ off-peak mode together with the first
three peak modes. The extracted amplitudes are displayed in the lower-right
panel, where an amplitude plateau for all four modes appears within the range
$12\lesssim\left(  t_{0}-t_{peak}\right)  /M\lesssim16$. Notably, the
amplitude of the $n_{0}=0$ off-peak mode is approximately $\mathcal{O}\left(
10^{-4}\right)  $ smaller than those of the peak modes, further confirming
that the off-peak modes make a negligible contribution to the time-domain signal.

\begin{figure*}[ptb]
\includegraphics[width=0.95\linewidth]{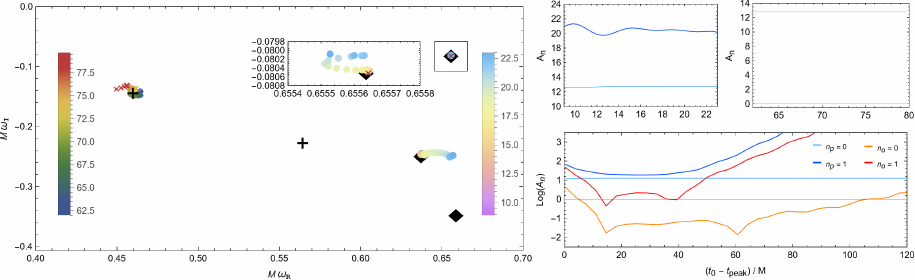}\caption{QNM fitting
results for $l=2$ with $Q/M=1.03507$. The waveform is fitted using the strong
fit model with two QNMs and the weak fit model with $\left(  N_{p}%
,N_{o}\right)  =\left(  2,2\right)  $. \textbf{Left}: QNM frequencies
extracted from the strong fit model in the complex plane. Dots and crosses
correspond to early and late fitting windows, respectively. For early windows,
only the peak modes are recovered, indicating that the early-time waveform is
dominated by these modes. For late windows, both the $n_{p}=0$ peak and
$n_{o}=0$ off-peak modes are extracted, as they are the slowest-decaying QNMs.
\textbf{Upper right}: Extracted amplitudes $A_{n}$ from the strong fit model
as functions of the start time $t_{0}$ for early and late fitting windows.
\textbf{Lower right}: Amplitudes obtained from the weak fit model. The
off-peak mode amplitudes are approximately $\mathcal{O}\left(  10^{-3}\right)
-\mathcal{O}\left(  10^{-1}\right)  $ smaller than those of the peak modes,
confirming their subdominant contribution to the time-domain waveform.}%
\label{fig: fitl2}%
\end{figure*}

Fig. \ref{fig: fitl2} presents the fitting results for the case $l=2$ with
$Q/M=1.03507$, where the waveform is analyzed using the strong fit model with
two QNMs and the weak fit model with $\left(  N_{p},N_{o}\right)  =\left(
2,2\right)  $. The left panel shows the QNM frequencies extracted from the
strong fit model in the complex plane, with dots and crosses representing
results obtained from early time windows $9\lesssim\left(  t-t_{peak}\right)
/M\lesssim23$ and late time windows $60\lesssim\left(  t-t_{peak}\right)
/M\lesssim80$, respectively. For early time windows, only the peak modes are
successfully extracted, indicating that the early portion of the waveform is
dominated by these modes. Using the weak fit model with the first two peak and
off-peak modes, we find that the extracted amplitudes exhibit a clear plateau
around $\left(  t-t_{peak}\right)  /M\sim25$, as shown in the lower-right
panel. The amplitudes of the $n_{0}=0$ and $n_{0}=1$ off-peak modes are
approximately $\mathcal{O}\left(  10^{-3}\right)  $ and $\mathcal{O}\left(
10^{-1}\right)  $ smaller than those of the peak modes, respectively. For late
time windows, the strong fit model captures both the $n_{p}=0$ peak mode and
the $n_{o}=0$ off-peak mode, which are the two slowest-decaying QNMs and thus
dominate the late-time waveform. The upper-right panels further show that the
$n_{p}=0$ peak mode provides the most significant contribution to the
late-time signal.

\section*{\textbf{Conclusions}}

In this work, we have investigated the QNM spectrum of a test scalar field in
the background of hairy BHs, using both frequency- and time-domain analyses.
As the BH charge-to-mass ratio increases, the scalar effective potential
transitions from a single-peaked to a double-peaked structure. When a
secondary potential barrier emerges, the additional characteristic scale
associated with the potential valley between the two barriers can destabilize
the QNM spectrum. By employing spectral methods, we computed the QNM spectrum
across both regimes and uncovered a remarkable phenomenon: even within the
single-peaked regime, where no explicit potential well exists, the residual
influence of this additional scale persists. This residual effect gives rise
to a spectrally unstable family of modes, referred to as off-peak modes, as
they are localized away from the potential peak. In contrast, the familiar
peak modes, situated near the potential peak, remain considerably more
spectrally stable.

We subsequently examined QNM extraction from time-domain signals in the
high-multipole case with $l=10$. In this case, the peak modes maintain strong
spectral stability even when the fundamental QNM is overtaken and destabilized
by the off-peak family. Through time-domain extractions at the overtaking
point, we found that the spectrally unstable off-peak modes contribute
negligibly to the waveform and are difficult to extract from early-time
signals. Although only the overtaking case was presented in this work, we also
performed additional extractions in parameter regions beyond the overtaking
point, approaching the double-peaked regime. These additional analyses led to
the same conclusion, providing a concrete example that the spectral
instability of the fundamental QNM does not necessarily imply instability in
the physical information extracted from early-time ringdown signals.

We also analyzed the low-multipole case with $l=2$, which is not only more
relevant to gravitational-wave observations but also of theoretical interest.
In this case, the single-peaked effective potential is more susceptible to
residual influences from a double-peaked structure, rendering the peak modes
less spectrally stable. This setup offers an ideal opportunity to explore how
the relative spectral stability between the two QNM families affects their
extraction from time-domain signals. We found that as the spectral stability
gap between the two families narrows, the contribution from the less stable
family becomes more significant. Interestingly, this enhanced contribution
allows us to successfully extract one off-peak mode at late times, providing
direct time-domain evidence for the existence of off-peak modes.

The main conclusion of this work is that when two distinct QNM families
coexist, their relative spectral stability plays a critical role in shaping
the prompt ringdown signal. The family exhibiting greater spectral stability
tends to dominate the time-domain waveform, while the less stable family
contributes only subdominant features. Moreover, as the difference in spectral
stability between the two families increases, the contribution from the less
stable modes becomes progressively suppressed.

Our findings have important implications for the BH spectroscopy program. By
revealing that the prompt ringdown is predominantly governed by the more
spectrally stable QNM family, we demonstrate that the modes most accessible to
observation are also those least sensitive to small perturbations in the
background spacetime. This connection between spectral stability and
observational dominance provides a natural self-selection mechanism that
enhances the robustness of BH spectroscopy: even when multiple QNM families
coexist, the gravitational-wave signal is expected to be dominated by the
spectrally stable modes, ensuring that the extracted frequencies remain
reliable and physically meaningful.

Finally, having analyzed spectral instability within a physically viable model
using a test scalar field, a natural next step is to extend this investigation
to the full perturbation spectrum of the BH, including gravitational,
electromagnetic, and scalar perturbations. Such an extension, however,
presents significant challenges: the full perturbation equations form a
coupled system of differential equations, and computing higher-overtone QNMs
on numerically constructed BH backgrounds is technically demanding
\cite{Myung:2018jvi,Blazquez-Salcedo:2020jee,Melis:2024kfr}. We leave a
detailed exploration of these coupled perturbations and their spectral
properties for future work.

\begin{acknowledgments}
We are grateful to Yiqian Chen and Lang Cheng for useful discussions and
valuable comments. This work is supported in part by NSFC (Grant No. 12275183
and 12275184).
\end{acknowledgments}

\bibliographystyle{unsrturl}
\bibliography{ref}

\section*{Supplemental Material}

In the Supplemental Material, we provide additional details and evidence
supporting several key physical statements presented in the main text.

\subsection{Spectral Method}

In this work, we employ spectral methods to numerically construct the hairy BH
solutions and compute their QNM spectra. Spectral methods are a
well-established technique for solving differential equations
\cite{boyd2001chebyshev}, in which the solution is approximated by a finite
linear combination of basis functions. A key advantage of spectral methods
lies in their exponential convergence rate for smooth functions, as opposed to
the linear or polynomial convergence typically achieved by finite difference
or finite element methods.

\subsubsection{BH Solution}

To construct a static, spherically symmetric BH solution, we adopt the ansatz
\begin{align}
ds^{2}  &  =-N(r)e^{-2\delta(r)}dt^{2}+\frac{1}{N(r)}dr^{2}+r^{2}%
d\Omega,\nonumber\\
A_{\mu}dx^{\mu}  &  =\Phi(r)dt\text{ and}\ \phi=\phi(r), \label{eq:HBH}%
\end{align}
which leads to the following equations of motion,
\begin{align}
N^{\prime}(r)  &  =\frac{1-N(r)}{r}-\frac{Q^{2}}{r^{3}e^{\alpha\phi^{2}(r)}%
}-rN(r)\left[  \phi^{\prime}(r)\right]  ^{2},\nonumber\\
\left[  r^{2}N(r)\phi^{\prime}(r)\right]  ^{\prime}  &  =-\frac{\alpha
Q^{2}\phi(r)}{r^{2}e^{\alpha\phi^{2}(r)}}-r^{3}N(r)\left[  \phi^{\prime
}(r)\right]  ^{3},\nonumber\\
\delta^{\prime}(r)  &  =-r\left[  \phi^{\prime}(r)\right]  ^{2},\label{eq:EOM}%
\\
\Phi^{\prime}(r)  &  =\frac{Q}{r^{2}e^{\alpha\phi^{2}(r)}}e^{-\delta
(r)},\nonumber
\end{align}
where the integration constant $Q$ denotes the BH electric charge, and primes
represent derivatives with respect to $r$. The boundary conditions are
determined by requiring regularity at the event horizon and asymptotic
flatness at spatial infinity.

For the numerical implementation, we introduce a compactified radial
coordinate defined by
\begin{equation}
y=1-\frac{2r_{H}}{r},
\end{equation}
which maps the event horizon at $r=r_{H}$ and spatial infinity to $y=-1$ and
$y=1$, respectively. In terms of this new coordinate, the relevant
functions---collectively denoted by $\mathcal{F}=\left\{  N,\delta,\Phi
,\phi\right\}  $---are expanded in a Chebyshev spectral series,
\begin{equation}
\mathcal{F}^{\left(  k\right)  }=%
%TCIMACRO{\dsum \limits_{i=0}^{N_{y}-1}}%
%BeginExpansion
{\displaystyle\sum\limits_{i=0}^{N_{y}-1}}
%EndExpansion
\alpha_{i}^{\left(  k\right)  }T_{i}\left(  x\right)  , \label{eq:sexpansion}%
\end{equation}
where $N_{y}$ is the number of collocation points, $T_{i}\left(  x\right)  $
denotes the Chebyshev polynomials, and $\alpha_{i}^{\left(  k\right)  }$ are
the spectral coefficients. Substituting the spectral expansion into Eq.
$\left(  \ref{eq:EOM}\right)  $ and evaluating the resulting expressions at
the Gauss-Chebyshev collocation points yield a system of nonlinear algebraic
equations for $\alpha_{i}^{\left(  k\right)  }$. These equations are solved
iteratively using the Newton-Raphson method, with each linearized system
efficiently handled by the built-in LinearSolve function in Mathematica.

We have also computed the BH solutions using the commonly adopted shooting
method and found excellent agreement with those obtained via the spectral
method. In this work, however, we employ the spectral method to construct the
BH solutions, as it provides superior accuracy near the boundaries. This
advantage is particularly important for subsequent QNM calculations, where
high-resolution accuracy and well-behaved boundary conditions are essential.

\subsubsection{QNM}

In the frequency domain, QNM frequencies are determined by solving Eq.
$\left(  \ref{eq:PertubaEq}\right)  $ for $\omega$ and $\tilde{\psi}\left(
\omega,x\right)  $. For a given QNM frequency $\omega$, the scalar field
perturbation $\tilde{\psi}\left(  \omega,x\right)  $ is required to be purely
ingoing at the BH horizon and purely outgoing at spatial infinity, as dictated
by causality. By analyzing the asymptotic behavior of Eq. $\left(
\ref{eq:PertubaEq}\right)  $, we find that $\tilde{\psi}\left(  \omega
,x\right)  $ behaves as
\begin{align}
\tilde{\psi}\left(  \omega,x\right)   &  \sim\left(  1-r/r_{H}\right)
^{-i\omega/\sqrt{f_{H}h_{H}}}\text{, }r\rightarrow r_{H}\text{,}\nonumber\\
\tilde{\psi}\left(  \omega,x\right)   &  \sim e^{i\omega r}\left(
r/r_{H}\right)  ^{-i\left(  f_{I}+h_{I}\right)  \omega/2}\text{, }%
r\rightarrow\infty,
\end{align}
where the coefficients $f_{H}$, $h_{H}$, $f_{I}$, and $h_{I}$ are determined
from the asymptotic expansions of the metric functions near the horizon and at
spatial infinity,%
\begin{align}
N(r)e^{-2\delta(r)}  &  \sim f_{H}\left(  r-r_{H}\right)  \text{, }N(r)\sim
h_{H}\left(  r-r_{H}\right)  ,\nonumber\\
N(r)e^{-2\delta(r)}  &  \sim1+\frac{f_{I}}{r}\text{, }N(r)\sim1+\frac{h_{I}%
}{r}.
\end{align}
It is evident that for a stable QNM, whose frequency $\omega$ has a negative
imaginary part, the corresponding perturbation $\tilde{\psi}\left(
\omega,x\right)  $ diverges both at the horizon and at spatial infinity.

A key step in numerically solving Eq. $\left(  \ref{eq:PertubaEq}\right)  $
for QNM frequencies using spectral methods is to remove the divergences in
$\tilde{\psi}\left(  \omega,x\right)  $. This can be achieved either by
factoring out the asymptotic behavior of $\tilde{\psi}\left(  \omega,x\right)
$ or by introducing hyperboloidal coordinates \cite{Zenginoglu:2007jw,Zenginoglu:2011jz,PanossoMacedo:2023qzp}. In
the former approach, we define a new function $u\left(  x\right)  $ through
the following decomposition,%
\begin{equation}
\tilde{\psi}\left(  \omega,x\right)  =e^{i\omega r}\left(  \frac{r}{r_{H}%
}\right)  ^{-i\frac{\left(  f_{I}+h_{I}\right)  \omega}{2}}\left(  1-\frac
{r}{r_{H}}\right)  ^{-\frac{i\omega}{\sqrt{f_{H}h_{H}}}}u\left(  x\right)
\text{,}%
\end{equation}
which, when substituted into Eq. $\left(  \ref{eq:PertubaEq}\right)  $, yields
the differential equation governing $u\left(  x\right)  $. In the latter
approach, we introduce compact hyperboloidal coordinates $\left(
\tau,y\right)  $ via the transformation,%
\begin{equation}
t=r_{H}\left[  \tau-H\left(  y\right)  \right]  \text{, }r=\frac{2r_{H}}{1-y},
\label{eq:coorTrans}%
\end{equation}
where the height function $H\left(  y\right)  $ is defined as%
\begin{equation}
H\left(  y\right)  =\frac{e^{-\delta\left(  -1\right)  }\ln\left(  1+y\right)
}{n\left(  -1\right)  }+\frac{2}{1-y}+\ln\left(  1-y\right)  \left[
1+2n^{\prime}\left(  1\right)  \right]  .
\end{equation}
Here, $n\left(  y\right)  =2N\left(  y\right)  /\left(  1+y\right)  $.
Performing a Fourier transformation with respect to $\tau$,
\begin{equation}
\psi=\int d\omega u\left(  y\right)  e^{-i\omega\tau}\text{,}%
\end{equation}
we obtain the differential equation for $u\left(  y\right)  $ from Eq.
$\left(  \ref{eq:t-x eq}\right)  $. Since the constant-$\tau$ hypersurfaces
smoothly penetrate both the horizon and future null infinity, the transformed
field $u\left(  y\right)  $ remains finite as $x\rightarrow\pm1$.

In both approaches, the QNM equation takes the general form%
\begin{equation}
c_{0}\left(  y,\omega\right)  u\left(  y\right)  +c_{1}\left(  y,\omega
\right)  u^{\prime}\left(  y\right)  +c_{2}\left(  y,\omega\right)
u^{\prime\prime}\left(  y\right)  =0,
\end{equation}
where each coefficient $c_{i}$ is at most quadratic in $\omega$. To solve this
equation numerically, we discretize it at the Gauss-Chebyshev collocation
points, replacing derivatives with the corresponding spectral derivative
matrices. The discretized equation can then be written as
\begin{equation}
\left(  M_{0}+\omega M_{1}+\omega^{2}M_{2}\right)  u=0,
\end{equation}
where $M_{i}$ are purely numerical matrices. This generalized eigenvalue
problem is solved using Mathematica's built-in function Eigenvalues after
suitable algebraic manipulation \cite{Jansen:2017oag}. To minimize
interpolation errors, both the BH background and the QNMs are computed on the
same spectral grid, with the matrices $M_{i}$ evaluated directly from the
metric functions at the grid points. The QNMs obtained from the peeling-off
and hyperboloidal approaches show excellent agreement, confirming the
consistency of our numerical implementation.

To assess the numerical error and convergence behavior, we compute the QNM
frequencies for a sequence of increasing spectral resolutions $N_{y}$ with a
fixed step size $\Delta N_{y}$. The numerical error of the QNM frequency
$\omega_{N_{y}}$ obtained at resolution $N_{y}$ is estimated as%
\begin{equation}
\epsilon_{N_{y}}\equiv\left\vert \omega_{N_{y}}-\omega_{N_{y}-\Delta N_{y}%
}\right\vert +\left\vert \omega_{N_{y}}-\omega_{N_{y}+\Delta N_{y}}\right\vert
.
\end{equation}
For an exponentially convergent spectral expansion, $\epsilon_{N_{y}}$
decreases approximately exponentially with increasing $N_{y}$ until reaching a
plateau. In this work, we vary $N_{y}$ from $30$ to $140$ with a step of
$\Delta N_{y}=10$ and adopt the smallest $\epsilon_{N_{y}}$ as an estimate of
the numerical error of the QNM frequencies. Only QNM frequencies that exhibit
exponential convergence and possess a numerical error below $10^{-5}$ are
included in our analysis.

\subsection{Migration Rate of QNM}

\begin{figure*}[ptb]
\includegraphics[scale=0.95]{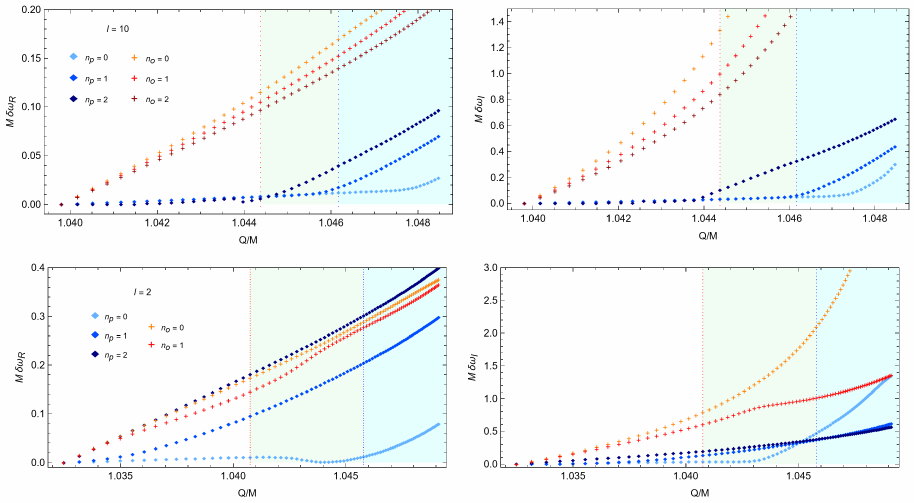}\caption{Migration rates $\delta\omega
_{R}$ (\textbf{Left}) and $\delta\omega_{I}$ (\textbf{Right}) for the real and
imaginary parts of the QNM frequencies, shown as a function of $Q/M$. The
upper and lower rows correspond to the $l=10$ and $l=2$ cases, respectively.}%
\end{figure*}

For a discrete series of QNMs $\left\{  \omega_{1},\cdots,\omega_{n}%
,\cdots,\omega_{N}\right\}  $ evaluated at corresponding discrete parameter
values, the migration of the real and imaginary parts of $\omega_{n}$ can be
quantified using the average normalized local migration rate over the interval
$\left[  \omega_{1},\omega_{n}\right]  $,%
\begin{equation}
\delta\omega_{R,I}=\sum\limits_{i=1}^{n-1}\frac{\left\vert \omega_{i+1}%
^{R,I}-\omega_{i}^{R,I}\right\vert }{\omega_{i}},
\end{equation}
where the indices $R$ and $I$ denote the real and imaginary parts,
respectively. In the continuum limit, $\delta\omega_{R,I}$ reduces to
$\ln\left\vert \omega_{n}^{R,I}/\omega_{1}^{R,I}\right\vert $.

Fig. 5 shows the migration rates $\delta\omega_{R,I}$ for the $l=10$ and $l=2$
cases in the upper and lower rows, respectively, where $\omega_{1}$ and
$\omega_{N}$ correspond to the QNMs at the smallest and largest $Q/M$. In the
$l=10$ case, the migration rates of the peak modes are significantly smaller
than those of the off-peak modes. In the $l=2$ case, the imaginary parts of
the QNMs exhibit noticeably lower migration rates for the peak modes in the
single-peaked regime. Around $Q/M=1.035$, the real parts of the $n_{p}=0$ and
$n_{p}=1$ peak modes also show reduced migration rates. Nevertheless, the
$n_{p}=2$ peak mode exhibits a migration rate $\delta\omega_{R}$ comparable to
that of the off-peak modes across the entire parameter range.

\subsection{Time Evolution of Waveforms}

\begin{figure*}[ptb]
\includegraphics[scale=0.75]{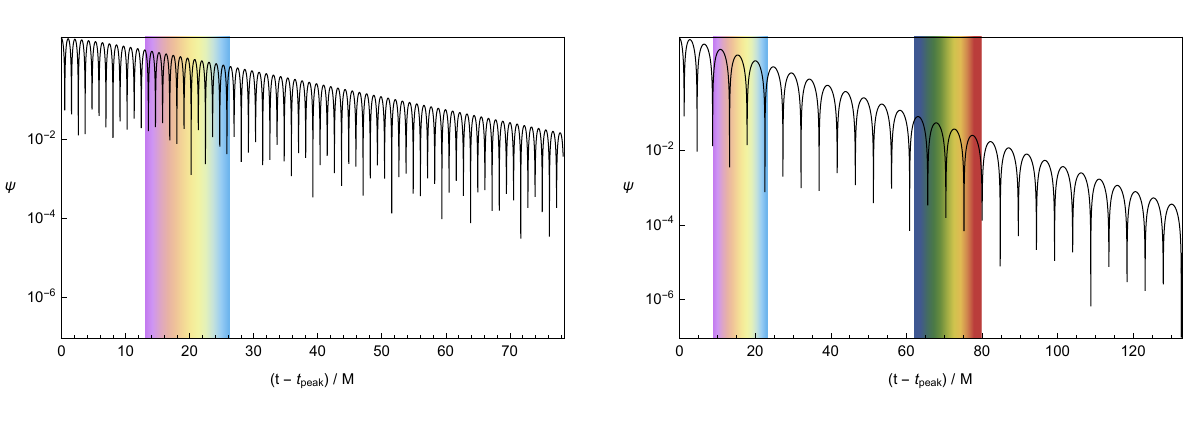}\caption{The left and right panels
correspond to the waveforms analyzed in Figs. \ref{fig: fitl10} and
\ref{fig: fitl2}, respectively. The start times of the time windows used for
the strong fit model extractions are indicated by consistent colormaps.}%
\label{fig:waveform}%
\end{figure*}

To compute the time-domain evolution of the scalar field waveform, we employ a
finite-difference scheme to discretize the wave equation $\left(
\ref{eq:t-x eq}\right)  $ \cite{Zhu:2014sya,Huang:2015cha}. Denoting
$\psi_{i,j}=\psi\left(  i\Delta t,j\Delta x\right)  $ and $V_{j}%
=V_{\text{eff}}\left(  j\Delta x\right)  $, the field evolution is given by
\begin{align}
\psi_{i+1,j} &  =-\psi_{i-1,j}+\frac{\Delta t^{2}}{\Delta x^{2}}\left(
\psi_{i,j+1}+\psi_{i,j-1}\right)  \nonumber\\
&  +\left(  2-2\frac{\Delta t^{2}}{\Delta x^{2}}-\Delta t^{2}V_{j}\right)
\psi_{i,j},\label{eq:disc}%
\end{align}
where we set $\Delta t/\Delta x=1/2$ to ensure numerical stability
\cite{Konoplya:2013rxa}. The initial condition is a Gaussian pulse,
$\psi\left(  t=0,x\right)  =\exp\left[  -\left(  x-a\right)  ^{2}/2\sigma
^{2}\right]  $, with $\psi\left(  t<0,x\right)  =0$, where $a$ is chosen near
the potential peak. To validate this discretization scheme, we also solve the
wave equation using Mathematica's built-in function NDSolve with both
finite-difference and finite-element methods, finding excellent agreement.
FIG. \ref{fig:waveform} displays the resulting waveforms, from which QNMs are
extracted in Figs. \ref{fig: fitl10} and \ref{fig: fitl2}. The start times of
the time windows used for the strong fit model extractions are indicated using
consistent colormaps.

\subsection{Weak Fit Model}

In this model, the QNM frequencies are fixed a priori, and only the amplitudes
and phases $\left\{  A_{n},\phi_{n}\right\}  $ are extracted, resulting in a
total of $2\times N$ fitting parameters for $N$ QNMs. To linearize the fitting
problem, the model is rewritten as
\begin{equation}
\psi\left(  t\right)  ={\sum_{n=0}^{N-1}}e^{\mathrm{\,Im}\,\omega_{n}t}\left[
C_{n}\sin\left(  \mathrm{Re}\,\omega_{n}\right)  +D_{n}\cos\left(
\mathrm{Re}\,\omega_{n}\right)  \right]  ,
\end{equation}
where $C_{n}=A_{n}\cos\phi_{n}$, $D_{n}=A_{n}\sin\phi_{n}.$ We perform a
linear fit using Mathematica's built-in function LinearModelFit within a
selected time window to extract $\left\{  C_{n},D_{n}\right\}  $, from which
the amplitudes and phases are reconstructed as
\begin{equation}
A_{n}=\sqrt{C_{n}^{2}+S_{n}^{2}},\,\phi_{n}=\arctan\left(  D_{n}/C_{n}\right)
.
\end{equation}
A plateau region typically emerges across the simultaneously fitted
parameters, indicating a reliable extraction
\cite{Baibhav:2023clw,Takahashi:2023tkb}. In the fitting procedure, we include
as many QNMs as possible, provided that a plateau region is observed.

\end{document}